\documentstyle[preprint,aps,epsf]{revtex}                 
\input epsf
\tightenlines
\begin{document}
\pagestyle{empty}                                      
\preprint{
\font\fortssbx=cmssbx10 scaled \magstep2
\hbox to \hsize{
\hfill$\raise .5cm\vtop{              
\hbox{NSC-NCTS-870901}}$}
}
\draft
\vfill
\title{ 
$B \to \eta' X_s$ in the Standard Model}

\vfill
\author{Xiao-Gang He$^{ab}$\thanks{e-mail address: hexg@phys.ntu.edu.tw} 
and Guey-Lin Lin$^c$
\thanks{e-mail address: glin@beauty.phys.nctu.edu.tw}}
\address{$^a$
\rm  Department of Physics, National Taiwan University,
Taipei, Taiwan, 10764, R.O.C.
}
\address{$^b$
\rm School of Physics, University of Melbourne, Parkville, Vic. 3052,
Australia.
}
\address{$^c$
\rm Institute of Physics, National Chiao-Tung University,
Hsinchu, Taiwan, 300, R.O.C.
}
%
%
\vfill
\maketitle
\begin{abstract}
We study $B \to \eta' X_s$ within the framework of the Standard  
Model.
Several mechanisms
such as $b \to \eta' s g$ through the QCD anomaly, and 
$b \to \eta's $ and $B\to \eta' s \bar q$ arising from four-quark  
operators
are treated simultaneously.
Using QCD equations of motion, we relate the effective Hamiltonian 
for the first mechanism to that for the latter
two. 
By incorporating next-to-leading-logarithmic(NLL) contributions,
the first mechanism is shown to give a significant branching ratio 
for $B\to \eta' X_s$, while the other 
two mechanisms account for about 15\% of the experimental value. 
The Standard Model
prediction for $B\to \eta' X_s$ is consistent with the CLEO data.    
\end{abstract}
%
%
%
%
\pagestyle{plain}


The recent observation of $B\to \eta' K$\cite{CLEO1} and 
$B\to \eta^{\prime}X_s$\cite{CLEO2} decays with high momentum
$\eta^{\prime}$ mesons has stimulated many theoretical activities
\cite{AS,HT,he1,KP,FR,excl,excl1,HZ}. 
One of the mechanisms proposed to account for this decay is
$b\to sg^*\to
sg\eta^{\prime}$\cite{AS,HT} where the $\eta^{\prime}$ meson is  
produced 
via the anomalous 
$\eta'-g-g$ coupling. According to a previous analysis\cite{HT},
this mechanism within the Standard Model(SM) can only account for 1/3 
of the 
measured branching ratio: ${\cal B}(B\to \eta^{\prime}X_s) =
\left[6.2\pm 1.6(\rm stat)\pm 1.3(\rm syst)^{+0.0}_{-1.5}(\rm bkg)
\right]\times 10^{-4}$\cite{CLEO2} with $2.0 < p_{\eta'} < 2.7$ GeV. 
There are also other calculations of $B \to \eta' X_s$ 
based on four-quark operators of the effective  
weak-Hamiltonian\cite{he1,KP}. 
These contributions to the branching ratio, typically $10^{-4}$, 
are also too small to account for $B\to \eta' X_s$, 
although the four-quark-operator
contribution is capable of explaining the branching ratio for
the exclusive $B \to \eta' K$ decays\cite{excl,excl1}.
These results
have inspired proposals for an enhanced $b\to sg$ and other 
mechanisms arising from physics 
beyond the Standard Model\cite{HT,KP,FR}. 
In order to see if new physics should play any role in $B\to \eta' X_s$,
one has to have a better understanding on the SM prediction.
In this letter, we carry out a careful analysis on 
$B\to \eta' X_s$ in the SM 
using next-to-leading effective Hamiltonian and consider several
mechanisms simultaneously.

We have observed that all earlier 
calculations on $b\to sg\eta^{\prime}$ were either based upon  
one-loop
result\cite{HT} which neglects the running of QCD renormalization 
-scale from $M_W$ 
to $M_b$ or only taking into account part of the running  
effect\cite{AS}. 
Since the short-distance
QCD effect is generally significant in weak decays, it is 
therefore crucial to compute $b\to sg\eta^{\prime}$ using the 
effective Hamiltonian approach. As will be shown later, the process
$b\to sg\eta^{\prime}$ alone contribute significantly to 
$B\to \eta' X_s$ 
while contributions   
from $b \to \eta's$ and $B\to \eta' s \bar q$ are suppressed.

The effective Hamiltonian\cite{REVIEW} for the $B\to \eta'  
X_s$ decay is 
given by:
\begin{eqnarray}
H_{eff}(\Delta B=1)&&={G_F\over  
\sqrt{2}}[\sum_{f=u,c}V_{fb}V_{fs}^*(C_1(\mu)O_1^f(\mu)+C_2(\mu)O_2^f(\mu))
\nonumber\\
&&-V^*_{ts}V_{tb}(\sum_{i=3}^{6}C_i(\mu)O_i(\mu)
+C_8(\mu)O_8(\mu))],
\label{HAMI}
\end{eqnarray}
with\cite{LNO} 
\begin{eqnarray}
O_1^f &=& (\bar{s}_if_j)_{V-A}(\bar{f}_jb_i)_{V-A},\;\;
O_2^f = (\bar{s}_if_i)_{V-A}(\bar{f}_jb_j)_{V-A}  \nonumber \\
O_3 &=& (\bar{s}_ib_i)_{V-A}\sum_{q}(\bar{q}_jq_j)_{V-A},\;\;
O_4 = (\bar{s}_ib_j)_{V-A}\sum_{q}(\bar{q}_jq_i)_{V-A}  \nonumber
\\
O_5 &=& (\bar{s}_ib_i)_{V-A}\sum_{q}(\bar{q}_jq_j)_{V+A},\;\;
O_6 = (\bar{s}_ib_j)_{V-A}\sum_{q}(\bar{q}_jq_i)_{V+A}  \nonumber
\\
O_8 &=& -{g_s\over
4\pi^2}\bar{s}_i\sigma^{\mu\nu}(m_sP_L+m_bP_R)T^a_{ij}
b_jG^a_{\mu\nu},
\label{OPER}
\end{eqnarray}
where $V\pm A \equiv 1\pm\gamma_5$.
In the above, we have dropped $O_7$ since its contribution is  
negligible.
For numerical analyses, we use the scheme-independent 
Wilson coefficients discussed in Ref.\cite{BJLW,DH}.    
For $m_t = 175$ GeV, $\alpha_s (m_Z^2) = 0.118$ and $\mu = m_b = 5$  
GeV,
we have\cite{DH}
\begin{eqnarray}
C_1 &=& -0.313, \;\;C_2 = 1.150,\;\;C_3 = 0.017,\;\;C_4 = -0.037,\;\;
C_5= 0.010,\;\;C_6 = -0.045,\;\; 
\label{WILSON}
\end{eqnarray}
At the NLL level, the effective Hamiltonian is modified by one-loop
matrix elements which effectively change $C_i(\mu)$($i=3,\cdots ,6$) into 
$C_i(\mu)+\bar{C}_i(q^2,\mu)$ with
\begin{equation}
\bar{C}_4(q^2,\mu) = \bar{C}_6(q^2,\mu) = -3\bar{C}_3(q^2,\mu)=
-3\bar{C}_5(q^2,\mu)=-P_s(q^2,\mu),
\end{equation}
where
\begin{equation}
P_s(q^2,\mu) = {\alpha_s \over 8\pi} C_2(\mu) \left ( {10\over 9} + 
G(m_c^2,q^2,\mu)\right ),
\end{equation}
with
\begin{equation}
G(m_c^2,q^2,\mu)=
4\int x(1-x) \log \left (
{m_c^2 - x(1-x)q^2\over \mu^2}\right )dx.
\end{equation}
The coefficient $C_8$ is equal to $-0.144$ at $\mu=5 \ {\rm GeV}
$\cite{REVIEW}, and $m_c$ is taken to be $1.4$ GeV. 

Before we discuss the dominant  
$b\to sg\eta^{\prime}$ process, let us first work out  
the four-quark-operator contribution to 
$B\to \eta' X_s$ using the above effective Hamiltonian.
We follow the approach of Ref.\cite{AS,he1,he2} which uses factorization 
approximation to estimate various hadronic matrix elements. 
The four-quark operators
can induce three types of processes represented by
1) $<\eta'|\bar q \Gamma_1 b|B> <X_s|\bar s \Gamma_1'q|0>$,
2) $<\eta'|\bar q \Gamma_2 q|0><X_s|\bar s \Gamma b| B>$, and 
3) $<\eta' X_s|\bar s \Gamma_3q|0><0|\bar q\Gamma_3'b|B>$. 
Here $\Gamma^{(')}_i$ denotes appropriate gamma matirces. 
The contribution from 1) gives a ``three-body'' type of decay,
$B\to \eta' s \bar q$. The contribution from 2) gives a ``two-body''
type of decay $b\to s\eta'$.
The contribution from 3) is the annihilation
type which is relatively suppressed and will be neglected.
Note that there are inteferences between 1) and 2), so they must be coherently
added together\cite{he1}.

Several decay constants and form factors needed in the calculations  
are
listed below:
\begin{eqnarray}
&&<0|\bar u\gamma_\mu \gamma_5 u|\eta'> = 
<0|\bar d \gamma_\mu \gamma_5 d|\eta'>
=if_{\eta'}^u p^{\eta'}_\mu\nonumber\\
&&<0|\bar s \gamma_\mu \gamma_5 s|\eta'> = 
if_{\eta'}^s p^{\eta'}_\mu,\;\;
<0|\bar s \gamma_5 s|\eta'> = i(f_{\eta'}^u-f_{\eta'}^s) 
{m^2_{\eta'}\over 2m_s},\nonumber\\
&&f_{\eta'}^u = {1\over \sqrt{3}
} (f_1 \cos\theta_1 + 
{1\over \sqrt{2}} f_8 \sin\theta_8),\;\;
f_{\eta'}^s = {1\over \sqrt{3}}
(f_1\cos\theta_1 - \sqrt{2} f_8 \sin\theta_8),\nonumber\\
&&<\eta'|\bar u\gamma_\mu b|B^->=<\eta'|\bar d \gamma_\mu b|\bar B^0>
= F_1^{Bq}(p^B_\mu + p^{\eta'}_\mu) +
(F_0^{Bq}-F_1^{Bq}) {mB^2-m_{\eta'}^2\over 
q^2} q_\mu,\nonumber\\
&&F_{1,0}^{Bq}={1\over \sqrt{3}} ({1\over \sqrt{2}} 
\sin\theta F^{B\eta_8}_{1,0}
+\cos\theta F^{B\eta_1}_{1,0}).
\end{eqnarray}
Fot the $\eta'-\eta$ mixing associated with decay constants above,
we have used the two-angle 
-parametrization.
The numerical values of various parameters are obtained from 
Ref. \cite{FK} with 
$f_1= 157$ MeV, $f_8=168$ MeV,
and the mixing angles $\theta_1 = -9.1^0$, $\theta_8=-22.1^0$. 
For the mixing angle associated with form factors, we use the
one-angle parametrization with $\theta = - 15.4^o$\cite{FK},
since these form factors were calculated in that 
formulation\cite{he1,he2}. 
In the latter discussion of $b\to sg\eta'$,  
we shall use the same parametrization in order to compare our results with  
those of earlier works
\cite{AS,HT}.  
For form factors, we assume
that $F^{B\eta_1} = F^{B\eta_8} = F^{B\pi}$ with 
dipole and monopole $q^2$ dependence for $F_1$ and $F_0$,  
respectively.
We used the running mass $m_s \approx 120$ MeV
at $\mu = 2.5$ GeV and $F^{B\pi} = 0.33$ following Ref.\cite{excl1}.

The branching ratios of the above processes
also depend on two less well-determined KM matrix 
elements, $V_{ts}$ and $V_{ub}$. 
The dependences on $V_{ts}$ arise from the penguin-diagram contributions
while the dependences on $V_{ub}$ and its phase $\gamma$ occur through
the tree-diagram contributions. We will use $\gamma=64^0$ obtained from
Ref.\cite{PRS}, $\vert V_{ts}\vert \approx \vert V_{cb}\vert=0.038$
and $\vert V_{ub}\vert / \vert V_{cb}\vert =0.08$ for an illustration.
We find  
that, for $\mu=5$ GeV,
the branching ratio in the signal region $p_{\eta'} \geq 2.0$ GeV 
($m_X \leq 2.35$ GeV) is
\begin{eqnarray}
{\cal B}(b\to \eta' X_s) \approx 1.0 \times 10^{-4}.
\end{eqnarray}
The branching ratio can reach $2\times 10^{-4}$ if all parameters  
take 
values in favour of $B\to \eta' X_s$.
Clearly the mechanism by four-quark operator is not sufficient  
to explain the observed $B\to \eta^{\prime}X_s$ 
branching ratio.

We now turn to the major mechanism for $B\to \eta^{\prime}X_s$:
$b\to \eta' s g$ through the QCD anomaly. To see how the
effective Hamiltonian in Eq. (\ref{HAMI}) can be applied to calculate
this process,  
we rearrange part of the effective Hamiltonian such that
\begin{eqnarray}
\sum_{i=3}^6C_iO_i=(C_3+{C_4\over N_c})O_3
+(C_5+{C_6\over N_c})O_5
- 2(C_4-C_6)O_A+2(C_4+C_6)O_V,
\label{GLUE}
\end{eqnarray}
where 
\begin{equation}
O_A=\bar{s}\gamma_{\mu}(1-\gamma_5)T^a b \sum_{q}\bar{q}\gamma^{\mu}
\gamma_5T^a q,
\;\;O_V=\bar{s}\gamma_{\mu}(1-\gamma_5)T^a b \sum_{q}
\bar{q}\gamma^{\mu}
T^a q.
\end{equation} 
Since the light-quark bilinear in $O_V$ carries the quantum number
of a gluon, one expects\cite{AS} $O_V$ give contribution to
the $b\to sg^*$ form factors. In fact, by applying the QCD equation  
of motion
: $D_{\nu}G^{\mu\nu}_a=g_s\sum \bar{q}\gamma^{\mu}T^a q$,
we have $O_V=(1/ g_s)\bar{s}\gamma_{\mu}(1-\gamma_5)T^a b D_{\nu}
G^{\mu\nu}_a$\cite{EQM}. 
In this form, $O_V$ is easily seen to give rise to $b\to sg^*$ 
vertex.    
Let us write  
the effective $b\to sg^*$ vertex as
\begin{eqnarray}
\Gamma_{\mu}^{bsg}&=& -{G_F\over \sqrt{2}}  V_{ts}^*V_{tb}
{g_s\over 4\pi^2} (\Delta F_1 \bar s( q^2 \gamma_\mu - q\!\!\!/\  
q_\mu) LT^a b - i F_2 m_b \bar s \sigma_{\mu\nu}q^\nu RT^ab) .
\label{DECOM}
\end{eqnarray}
In the above, we define the form
factors $\Delta F_1$ and $F_2$ according to the convention in 
Ref. \cite{HT}. Inferring from Eq. (\ref{GLUE}), we arrive at 
\begin{eqnarray}
&&\Delta F_1 ={4\pi \over \alpha_s} (C_4(\mu)+C_6(\mu)),\;\;
F_2 =-2C_8(\mu)
\label{F12}
\end{eqnarray}
We note that our relative sign between $\Delta F_1$ and $F_2$ agree with 
those in Ref. \cite{HT,KP}, and shall
result in a destructive interference for the rate of $b\to sg\eta'$. 
We stress that this relative sign is fixed by treating
the sign of $O_8$ and the
convention of QCD covariant derivative consistently\cite{KA}. 
To ensure the sign, we
also check against the result by Simma and Wyler\cite{SW} on $b\to
sg^*$ form factors. An agreement on sign is found.     
Finally, we remark that, at the NLL level, 
$\Delta F_1$ should be corrected by one-loop matrix elements. The dominant
contribution arises from the operator $O_2$ where its charm-quark-pair 
meets to form a gluon. In fact, this contribution, denoted as 
$\Delta \bar{F}_1$ for convenience, has been 
shown in Eqs. (4)-(6), namely $\Delta \bar{F}_1
 ={4\pi \over \alpha_s} (\bar{C}_4(q^2,\mu)+\bar{C}_6(q^2,\mu))$.
 
To proceed further, we recall
the distribution of the 
$b(p)\to
s(p')+g(k)+\eta^{\prime}(k^{\prime})$ branching ratio\cite{HT}:
\begin{figure} [hbt] 
\epsfxsize=0pt
\centerline{\epsfbox{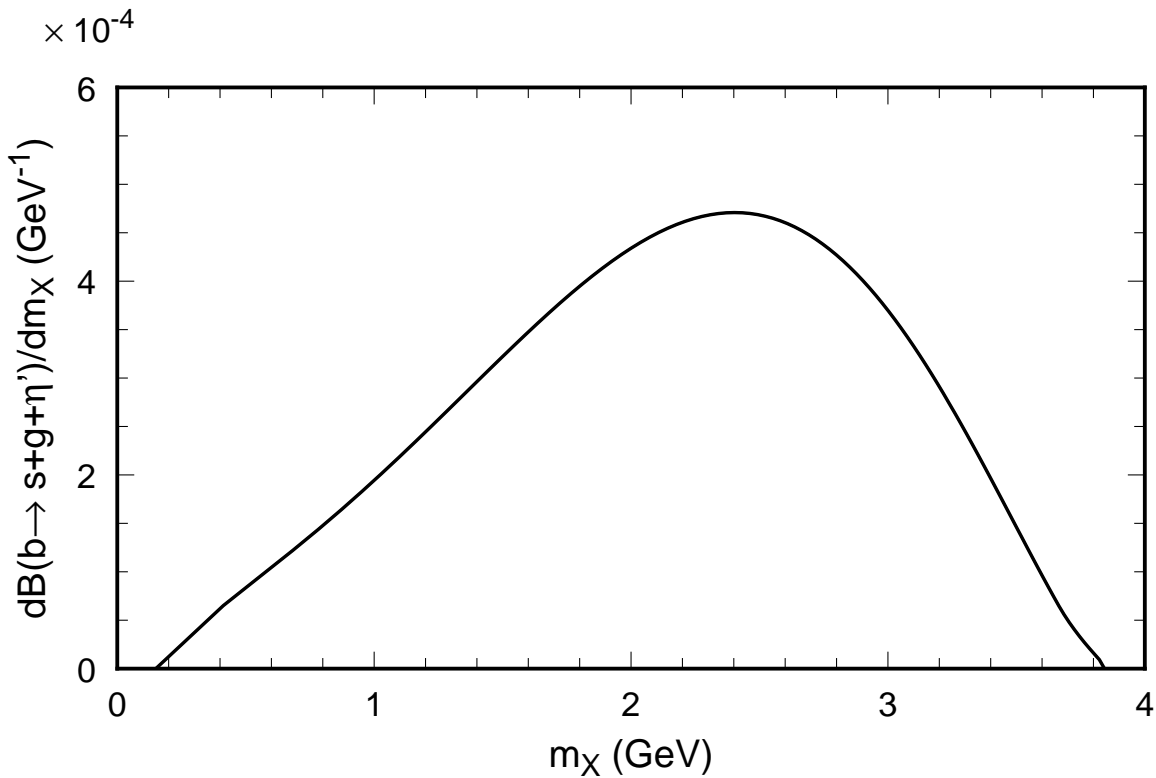}}
\vskip 1cm
\caption{The distribution of ${\cal B}(b\to s+g+\eta')$
as a function of the recoil mass 
$m_X$.}
\label{fig1}
\end{figure}
\begin{eqnarray} 
{d^2{\cal B}(b\to sg\eta')\over dx dy}
&\cong& 0.2\cos^2\theta
\left( {g_s(\mu)\over 4\pi^2} \right)^2 {a_g^2(\mu)m_b^2\over 4}
\nonumber \\
&\times& \left[ {\vert \Delta F_1\vert }^2 c_0 + Re(\Delta F_1 F_2^*)
{c_1\over y} + {\vert \Delta F_2\vert }^2 {c_2\over y^2} \right ],
\label{BR}
\end{eqnarray}
where $a_g(\mu)\equiv \sqrt{N_F}\alpha_s(\mu)/ \pi f_{\eta'}$ is 
the strength of $\eta'-g-g$ vertex: $a_g\cos\theta
\epsilon_{\mu\nu\alpha\beta}q^{\alpha}k^{\beta}$ with $q$ and $k$ the 
momenta of two gluons; 
$x\equiv (p^{\prime}+k)^2 / m_b^2$ and $y\equiv (k+k^{\prime})^2/  
m_b^2$; 
$c_0$, $c_1$ and $c_2$ are functions of $x$ and $y$ as 
given by:
\begin{eqnarray}
c_0 &=& \left[ -2x^2y+(1-y)(y-x^{\prime})(2x+y-x^{\prime})
\right ]/2, \nonumber \\ 
c_1 &=&(1-y)(y-x^{\prime})^2,\nonumber \\
c_2 &=& \left[ 2x^2y^2-(1-y)(y-x^{\prime})(2xy-y+x^{\prime})\right  
]/2, 
\end{eqnarray} 
with $x^{\prime}\equiv m_{\eta^{\prime}}^2 / m_b^2$; and the 
$\eta^{\prime}-\eta$ mixing angle
$\theta$ is taken to be $-15.4^o$ as noted earlier.
Finally, in obtaining the normalization factor: 0.2, we have taken into 
account the one-loop QCD correction\cite{SL} to the
semi-leptonic $b\to c$ decay for consistency.

In previous one-loop calculations without QCD corrections, 
it was found $\Delta F_1\approx -5$
and $F_2\approx 0.2$\cite{AS,HT}.
In our approach, we obtain $\Delta F_1=-4.86$ and $F_2=0.288$ 
from Eqs. (\ref{WILSON}) and (\ref{F12}). However, $\Delta F_1$ is enhanced 
significantly by the matrix-element correction
$\Delta \bar{F}_1(q^2,\mu)$. The latter quantity develops 
an imaginary part as $q^2$ passes the charm-pair threshold, 
and the magnitude of its real part also
becomes maximal at this threshold. From Eqs. (3), (4) and (5), one  
finds 
Re$(\Delta \bar{F}_1(4m_c^2,\mu))=-2.58$ at $\mu=5$ GeV.
Including the contribution by $\Delta \bar{F}_1(q^2,\mu)$ with $\mu=5$ GeV, 
and 
using Eq. (\ref{BR}), we find ${\cal B}(b\to sg\eta')=5.6\times  
10^{-4}$
with the cut $m_{X}\equiv \sqrt{(k+p')^2}\leq 2.35$ GeV imposed in 
the CLEO measurement\cite{CLEO2}. This branching ratio is 
consistent with CLEO's measurement on the $B\to \eta' X_s$
branching ratio\cite{CLEO2}.  
Without the kinematic cut, we obtain ${\cal B}(b\to sg\eta')=1.0\times  
10^{-3}$, which is much larger than $4.3\times 10^{-4}$ 
calculated previously\cite{HT}.
We also obtain the spectrum         
$d{\cal B}(b\to sg\eta')/dm_X$ as depicted in Fig. 1. 
The peak of the 
spectrum corresponds to $m_X\approx 2.4$ GeV. 

It is interesting to note that the CLEO analysis\cite{CLEO2}
indicates that, without the anomaly-induced contribution,
the recoil-mass($m_X$) spectrum of $B\to \eta' X_s$  can not be well 
reproduced even if the
four-quark operator contributions are normalized to
fit the branching ratio of the process.
On the other hand, if $b\to sg^*\to sg\eta'$ dominates the 
contributions to $B\to \eta' X_s$, as shown here,
the $m_X$ spectrum can be fitted better as shown in Fig. 2 of
Ref.\cite{CLEO2}. It is also interesting to remark that
although the four-quark operator contributions can not 
fit the branching ratio nor the spectrum, it does play a
role in producing a small peak in the spectrum, which
corresponds to the $B\to \eta'K$ mode. Specifically, the
$B\to \eta' K$ mode is accounted for by the $b\to s\eta'$ type
of decays discussed earlier. Based on results obtained so far,  
one concludes that
the Standard Model is not in conflict the experimental data
on $B\to \eta' X_s$. It can produce not only the branching
ratio for $B\to \eta'X_s$ but also the recoil-mass spectrum when
contributions from the anomaly mechanism and the four-quark
operators are properly treated.     
 
Up to this point, $a_g(\mu)$ of the  
$\eta'-g-g$ vertex has been treated as a constant independent 
of invariant-masses of the gluons, and $\mu$ is 
set to be $5$ GeV. In practice, $a_g(\mu)$ should behave like 
a form-factor which becomes suppressed as the gluons attached to it 
go farther off-shell\cite{AS,HT,KP}. However,
it remains unclear how much  
the form-factor suppression might be.
It is possible that the branching ratio 
we just obtained gets reduced significantly by the 
form-factor effect in $\eta'-g-g$ vertex.
Should a large form-factor suppression occur, the additional 
contribution from $b\to \eta' s$ and $B\to \eta' s \bar q$ 
discussed earlier would become crucial. 
We however like to stress that 
our estimate of $b\to sg\eta^{\prime}$
with $\alpha_s$ evaluated at $\mu=5$ GeV is conservative. To  
illustrate 
this, let us compare branching ratios for $b\to sg\eta^{\prime}$ 
obtained at $\mu=5$ GeV
and $\mu=2.5$ GeV respectively. In NDR scheme\cite{NDR}, 
branching ratios at the 
above two scales with the cut $m_X\leq 2.35$ GeV
are $4.9\times 10^{-4}$ and 
$9.1\times 10^{-4}$ respectively. One can clearly see the significant
scale-dependence! With the enhancement resulting from lowering the 
renormalization scale, there seems to be some room for the  
form-factor 
suppression in the attempt of explaining $B\to \eta^{\prime}X_s$ by
$b\to sg\eta^{\prime}$ 
\cite{FF}.  

It should be noted that the above scale-dependence
is solely due to the coupling constant 
$\alpha_s(\mu)$ appearing in the $\eta^{\prime}-g-g$
vertex. In fact, the $b\to sg^*$ vertex is rather insensitive
to the renormalization scale. Indeed, from Eq. (\ref{DECOM}),
we compute in the NDR scheme the scale-dependence of 
$g_s\cdot (\Delta F_1+\Delta \bar{F}_1(q^2))$. We find that, as
$\mu$ decreases from 5 GeV to 2.5 GeV, 
the peak value of the above
quantity increases by only $10\%$. Therefore, to stablize 
the scale-dependence, one should include corrections beyond those  
which
simply renormalize the $b\to sg^*$ vertex. We shall leave this to
a future investigation. 

It is instructive to compare our results with those of
Refs. \cite{AS,HT}. 
With the kinematic cut, our numerical result for ${\cal B}(b\to sg\eta')$ 
is only slightly smaller than the 
branching ratio, $8.2\times 10^{-4}$,  
reported in Ref. \cite{AS}, 
where the $\alpha_s(\mu)$ 
coupling of $\eta^{\prime}-g-g$ vertex is evaluated at $\mu \approx 1$ GeV,
and $\Delta F_1$ receives only short-distance contributions from
the Wilson coefficients $C_4$ and $C_6$. 
Although 
we have a much smaller $\alpha_s$, 
which is evaluatd at $\mu= 5$ GeV, 
and the interference of $\Delta F_1$ and $F_2$ 
is destructive\cite{HT} rather than constructive\cite{AS},
there exists a compensating enhancement
in $\Delta F_1$ due to one-loop matrix elements.
The branching ratio in 
Ref.\cite{HT} is $2-3$ times smaller than ours 
since it is given by a $\Delta F_1$ smaller than ours but 
comparable to that of   
Ref. \cite{AS}. 
Concerning the relative importance of 
$\Delta F_1$ and $F_2$, we find 
that $\Delta F_1$ alone gives 
${\cal B}(b\to sg\eta^{\prime})=6.5\times 10^{-4}$
with the kinematic cut $m_X\leq 2.35$ GeV.
Hence the inclusion of $F_2$ lowers down the branching ratio by only
$14\%$.  
Such a small interference effect is quite distinct from
results of Refs. \cite{AS,HT} where $20\%-50\%$ of 
interference effects are found. We attribute this to the enhancement of 
$\Delta F_1$ in our calculation.

Before closing we would like to comment on the branching ratio for
$B\to \eta X_s$. It is interesting to note that the width of $b\to \eta s g$  
is suppressed by $\tan^2\theta$ compared to that of 
$b\to\eta' sg$.
Taking $\theta=-15.4^o$, we obtain  
${\cal B}(B\to \eta X_s)\approx 4\times 10^{-5}$. The contribution 
from the four-quark operator can be  
larger. 
Depending on the choice of parameters, we find that $B(B\to \eta X_s)$
is in the range of $(6\sim 10)\times 10^{-5}$.

In conclusion, we have calculated the branching ratio of 
$b\to sg\eta^{\prime}$ by including the NLL correction to the $b\to  
sg^*$
vertex. By assuming a low-energy $\eta^{\prime}-g-g$ vertex, and
cutting the recoil-mass $m_X$ at $2.35$ GeV, we  
obtained
${\cal B}(b\to sg\eta^{\prime})=(5-9)\times 10^{-4}$ depending on the
choice of the QCD renormalization-scale. 
Although  
the form-factor suppression
in the $\eta^{\prime}-g-g$ vertex is anticipated, it remains possible  
that the anomaly-induced 
process $b\to sg\eta^{\prime}$ could account for the CLEO measurement
on ${\cal B}( B\to \eta^{\prime}X_s)$.      
For the four-quark operator contribution, we obtain
${\cal B}(B\to \eta' X_s) \approx 1\times 10^{-4}$. This accounts for roughly  
15\% of the
experimental central-value and can reach 30\% if favourable  
parameters are
used. 
Finally, combining contributions from the anomaly-mechanism and 
the four-quark operators, the entire range of $B\to \eta' X_s$ spectrum
can be well reproduced.

\acknowledgments
We thank W.-S. Hou, A. Kagan and A. Soni for discussions.
The work of XGH is supported by Australian Research 
Council and National Science Council of R.O.C. under the grant numbers
NSC 87-2811-M-002-046 and NSC 88-2112-M-002-041.
The work of GLL is supported by
National Science Council of R.O.C. under the grant numbers 
NSC 87-2112-M-009-038, NSC 88-2112-M-009-002, 
and National Center for Theoretcal Sciences of  
R.O.C. 
under the topical
program: PQCD, B and CP.   
%



\begin{references}
%
\bibitem{CLEO1}
CLEO collaboration, B.H. Behrens {\it et al.}, 
Phys. Rev. Lett. {\bf 80} (1998)  
3710.
%
\bibitem{CLEO2}
{CLEO Collaboration} T. E. Browder  {\it et al.},
Phys. Rev. Lett. {\bf 81} (1998) 1786.
%
\bibitem{AS}
D. Atwood and A. Soni, Phys. Lett. {\bf B 405} (1997) 150.
\bibitem{HT}
W.S. Hou and B. Tseng, Phys. Rev. Lett. {\bf 80} (1998) 434.
%
\bibitem{he1}
A. Datta, X.-G. He, and S. Pakvasa, Phys. Lett. {\bf B 419} (1998) 369.
%
\bibitem{KP}
A. L. Kagan and A. Petrov, Preprint hep-ph/9707354;
Preprint hep-ph/9806266.
%
\bibitem{FR}
%
H. Fritzsch, Phys. Lett. {\bf B 415} (1997) 83;
X.-G. He, W.-S. Hou and C.S. Huang, Phys. Lett. {\bf B 429} (1998) 99.
%
\bibitem{excl}
H.-Y. Cheng, and B. Tseng, Preprint hep-ph/9803457;
A. Ali, J. Chay, C. Greub and P. Ko, Phys. Lett. {\bf B424} (1998)  
161;
N. Deshpande, B. Dutta and S. Oh, Phys. Rev. {\bf D 57} (1998) 5723.
%
\bibitem{excl1} A. Ali, G. Kramer and C.-D. Lu, Preprint  
hep-ph/9804363.
%
\bibitem{HZ}
I. Halperin and A. Zhitnitsky, Phys. Rev. Lett. {\bf 80} (1998) 438;
F. Araki, M. Musakonov and H. Toki, Preprint hep-ph/9803356;
D.S. Du, Y.-D. Yang and G.-H. Zhu, Preprint hep-ph/9805451;
M. Ahmady, E. Kou and A. Sugamoto, Phys. Rev. {\bf D 58} (1998)  
014015;
D.S. Du, C.S. Kim and  Y.-D. Yang, Phys. Lett. {\bf B 426} (1998) 133;
F. Yuan and K.-T. Chao, Phys. Rev. {\bf D 56} (1997) 2495;
A. Dighe, M. Gronau and J. Rosner, Phys. Rev. Lett. {\bf 79} (1997)  
4333.
%
\bibitem{REVIEW}
For an extensive review on the subject of effective Hamiltonian, see 
G. Buchalla, A. J. Buras and
M. E. Lautenbacher, Review of Modern Physics,
{\bf 68} (1996) 1125, which contains a detailed list of original  
literatures. 
%
\bibitem{LNO}
The sign of $O_8$ is consistent with the 
covariant derivative, $D_{\mu}=\partial_{\mu}-igT^aA_{\mu}^a$, in 
the QCD Lagrangian. See,
A. Lenz, U. Nierste and Ostermaier, Phys. Rev. {\bf D 56} (1997) 7228.
%
\bibitem{BJLW}
A. Buras, M. Jamin, M. Lautenbacher and P.Weisz, Nucl. Phys. {\bf B  
370} (1992)
69.
%
\bibitem{DH}
N. G. Deshpande and X.-G. He, Phys. Lett. {\bf B 336} (1994) 471.
%
\bibitem{he2} T.E. Browder, A. Datta, X.-G. He and S. Pakvasa,
Phys. Rev. {\bf D 57} (1998) 6829.
%
\bibitem{FK} T. Feldmann, P. Kroll and B. Stech, e-print  
hep-ph/9802409.
%
\bibitem{PRS} F. Parodi, P. Roudeau and A. Stocchi, e-print hep-ph/9802289. 
%
\bibitem{EQM}
By applying 
the QCD equation
of motion or performing a direct calculation, it was shown that the operator
basis of $O_3-O_6$ are suitable to describe nonleptonic weak decays 
although effective vertices such as $s\to d+{\rm gluons}$ are encountered.
Here the operator basis on the r.h.s of Eq. (9) is more suitable for our 
purpose.
For detail, see A. I. Vainshtein {\it et al.}, JETP Lett. {\bf 22} (1975) 55; 
M. A. Shifman {\it et al.}, Nucl. Phys. {\bf B 120} (1977) 316;
M. B. Wise and E. Witten, Phys. Rev. {\bf D 20} (1979) 1216.
%
\bibitem{KA}
We thank A. Kagan for pointing out this to us, 
which helps us to detect a sign error in our earlier calculation.   
%
\bibitem{SW} H. Simma and D. Wyler, Nucl. Phys. {\bf B 344} (1990) 283.
%
\bibitem{SL} G. Corbo, Nucl. Phys. {\bf B 212} (1983) 99; 
N. Cabibbo, G. Corbo and L. Maiani, {\it ibid.}
{\bf B 155} (1979) 93.
%
\bibitem{NDR} 
In NDR scheme, apart from a different set of Wilson coefficients 
compared to Eq. (\ref{WILSON}), the constant term: ${10\over 9}$ 
at the r.h.s. of Eq. (5) is replaced by ${2\over 3}$. For details,
see, for example, Ali and Greub, Phys. Rev. 
{\bf D 57} (1998) 2996. 
%
\bibitem{FF}
We do notice that
$B(b\to sg\eta^{\prime})$ is suppressed by more than one order of magnitude  
if $a_g(\mu)$ in Eq. (\ref{BR}) is replaced by $a_g(m_{\eta'})\cdot  
{m_{\eta'}^2\over 
(m_{\eta'}^2-q^2)}$ according to Ref.\cite{KP}. 
However, this prescription for $a_g$ stems from the assumption that
$g^*\to g\eta'$ form factor behaves in the same way as the QED-anomaly
form factor $\gamma^*\to \gamma\pi^0$. It remains unclear 
as raised in Refs. \cite{AS,HT} that one could make such a connection 
between two distinct form factors. 
%
\end{references}
\end{document}